# Topologically steerable ultra-flat electro-optic combs via non-Hermitian frequency lattices

Weiwei Liu[1,§], Xiaolong Su[1,§], Bing Wang[1,*], Yongjie Wang[1], Pang Fu[1], Wenxian Zhang[1], Ruiqian Cheng[1], Songpan Hong[1], Qingbin Zhang[1,†], and Peixiang Lu[1,2,‡]

[1]*Wuhan National Laboratory for Optoelectronics and School of Physics, Huazhong University of Science and Technology, Wuhan 430074, China*

[2]*Hubei Key Laboratory of Optical Information and Pattern Recognition, Wuhan Institute of Technology, Wuhan 430205, China*

**ABSTRACT**. Electro-optic (EO) frequency combs convert microwave signals to the optical domain with deterministic repetition rates, enabling applications in optical telecommunications, microwave photonics, and LiDAR. However, generating flat EO combs has remained a long-standing challenge due to exponential decay of the spectral envelope inherent to bidirectional frequency coupling within conventional schemes. Here we reveal the inherent correspondence between flat-comb generation and the topological energy bands of non-Hermitian nonreciprocal frequency lattices, and experimentally realize ultra-flat combs with topologically controlled spectral flow. We derive the exact analytical condition for flat-comb generation, and find that it maps precisely onto the complex energy band, thereby establishing a direct connection between comb-envelope engineering and non-Hermitian topology. Experimentally, we observe an ultra-flat comb whose frequency evolution direction is governed by the topological winding and remains robust against detuning and phase fluctuations. This scheme can be further extended to long-range coupling and multi-lattice architectures, underscoring its versatility for flat dual- and multi-comb generation. Our work establishes a topologically steerable approach for achieving ultra-flat EO combs, paving the way for chip-scale spectral engineering via non-Hermitian topology.

***Introduction***—Electro-optic frequency combs (EOFCs) directly convert microwave signals into the optical domain with electronic tunability [1-4], making them indispensable for optical telecommunications [5,6], microwave photonics [7,8], and LiDAR [9,10]. In resonant EOFCs, modulation at an integer multiple of the cavity free spectral range (FSR) coherently enhances sideband generation, which leads to broadband comb formation with high spectral purity and low power consumption [3]. However, conventional resonant EOFCs suffer from a fundamental limitation that the bidirectional coupling under pure phase modulation forces a symmetric frequency quantum walk, causing the spectral envelope to decay exponentially with sideband order [11-13]. Considerable efforts have been devoted to achieving broadband and flat EO combs, including coupled-resonator designs [14], fast-gain quantum walk [15], and optical parametric oscillator [16]. Nevertheless, the intrinsic bidirectionality of frequency coupling has made the realization of spectrally flat EOFCs a long-standing challenge.

Recent advances in topological photonics have provided effective approaches for controlling wave dynamics in synthetic dimensions [17], giving rise to nontrivial phenomena such as topological energy bands [18-20], non-Hermitian skin effect [21-24], unidirectional amplification [25] and dynamic localization [26,27]. Synthetic frequency dimensions, where resonant modes map to discrete sites with controllable hoppings, provide an ideal platform for realizing topological physics [28-30]. Within this framework, growing technologies have been developed to flexibly regulate frequency evolution and comb generation. For instance, early theoretical and experimental efforts have utilized synthetic gauge fields and long-range coupling to tailor the comb spectral envelope [3], and non-Hermitian frequency lattices have been demonstrated to exhibit arbitrary topological windings [18] as well as asymmetric, chiral, or even localized spectral distributions [31,32]. However, whether non-Hermitian topology can be harnessed to achieve spectrally flat EOFCs and thereby confer a topological approach for spectral steering has remained unexplored.

In this work, we reveal the inherent correspondence between flat-comb generation and the non-Hermitian energy bands of frequency lattices with nonreciprocal hoppings, enabling topologically steerable ultra-flat EOFCs. The nonreciprocal hoppings enforce directional frequency transition and resolve the intrinsically bidirectional spectral expansion of conventional phase-modulated systems. Using temporal coupled-mode theory, we analyze the evolution dynamics and derive the exact condition for achieving perfect spectral flatness, which maps precisely onto trajectory of the complex non-Hermitian band. Experimentally, we demonstrate an ultra-flat comb with a highly uniform spectral envelope, and reveal that the spectral flow direction is governed by the topological winding of the complex band. The scheme can be further scaled to long-range coupling and multi-lattice configurations, underscoring its versatility for generating tunable and dual/multiple flat combs.

*Contact author: wangbing@hust.edu.cn
†Contact author: zhangqingbin@hust.edu.cn
‡Contact author: lupeixiang@hust.edu.cn
§These authors contributed equally to this work

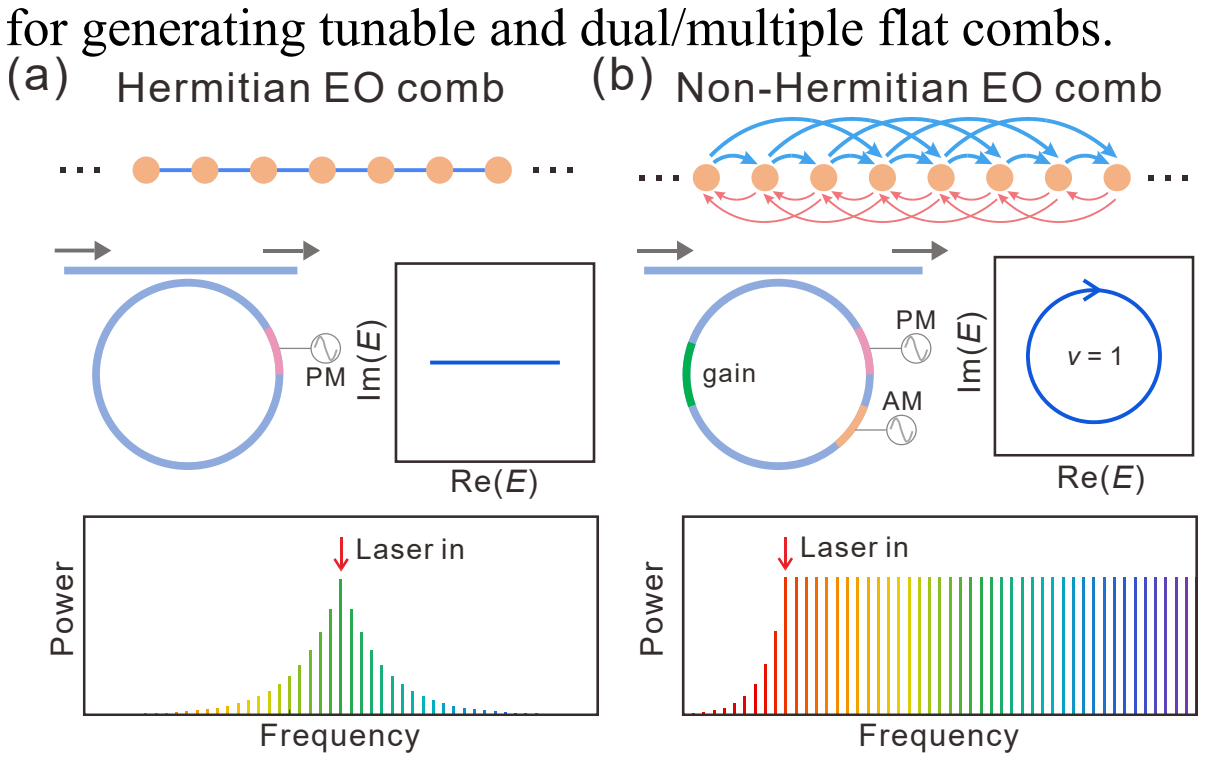

**Fig. 1. Principle of topologically steerable ultra-flat electro-optic (EO) frequency combs via non-Hermitian mechanism.** (a) Conventional Hermitian EO frequency comb based on phase modulation, which has a trivial topology and the spectrum shows exponential decay toward both higher- and lower-frequency sidebands. (b) Flat EO frequency comb enabled by non-Hermitian nonreciprocal coupling, which has a nontrivial topology and the spectrum can exhibit a perfectly flat envelope with directional flow.

***Theoretical model***—In a conventional resonant EOFC, a ring cavity is driven with a phase modulator (PM) to induce coherent transitions between adjacent frequency modes [Fig. 1(a)]. The reciprocal coupling can be described by a Hamiltonian $H_c = -\sum_m(Ja_m a^\dagger_{m+1} + Ja^\dagger_m a_{m+1})$, where $J$ denotes the coupling strength [29]. Consequently, an initially populated mode distribute energy to both higher and lower sidebands, producing an exponentially decaying spectral envelope. To break the reciprocity, we introduce an amplitude modulator (AM) into the cavity to establish a non-Hermitian landscape [18], creating an imbalance between forward and backward transitions [Fig. 1(b)]. The resulting nonreciprocal intermode coupling is described by $H'_c = -\sum_m(C_+ a_m a^\dagger_{m+n} + C_- a^\dagger_m a_{m+n})$, where $C_+$ and $C_-$ denote the forward and backward hopping amplitudes respectively. The nonreciprocal coupling takes the form $C_\pm = J \pm \Delta\exp(\pm i\varphi)$ ($\Delta \neq 0$), where $J$ is the reciprocal coupling strength provided by the PM, $\Delta$ is the non-Hermitian component induced by the AM, and $\varphi$ is the relative phase. While such non-Hermitian systems have been explored for exotic topological behaviors such as arbitrary topological windings [18] and complex-energy braiding [33], here we concentrate on their spectral characteristics for comb generation. By tailoring the nonreciprocal coupling ($C_+ \neq C_-$), we can achieve directional spectral transport along the frequency dimension, suppressing the bidirectional diffusion inherent to Hermitian lattices and enabling flat comb generation.

To quantitatively analyze the comb spectrum, we model the dynamics of frequency modes under continuous-wave (CW) excitation using temporal coupled-mode theory (TCMT) [18,28]. The evolution of the mode amplitude $a_m$ at resonant frequency $\omega_m = \omega_0 + m\Omega_R$ is governed by

$$\frac{\mathrm{d}a_m}{\mathrm{d}t} = -\frac{\gamma}{2}a_m + i\sum_{n\neq 0} C_n a_{m+n} + i\sqrt{\kappa}S_{in}e^{i\delta\omega t}\delta_{m,0}, \quad (1)$$

where $\gamma$ is the total cavity loss rate, and $\kappa$ is the waveguide-cavity coupling coefficient. $C_n$ represents the hopping amplitude from $(m+n)$-th to $m$-th modes. $S_{in}$ denotes the input field amplitude, and $\delta\omega$ is the laser-cavity detuning. The Kronecker symbol $\delta_{m,0}$ indicates that the input field primarily excites the central mode $a_0$ at frequency $\omega_0$. Under steady-state conditions, a perfectly flat comb corresponds to equal amplitudes of adjacent modes. Assuming $b_m = a_m e^{i\delta\omega t}$, and defining $L = b_m/b_{m-1} = e^{i\alpha}$, we obtain the flat-comb condition under steady-state operations (see Supplemental Material [34])

$$i\delta\omega + \frac{\gamma}{2} = i\sum_{n\neq 0} C_n e^{in\alpha} = -iE_{-\alpha}, \quad (2)$$

where $E_k = -\sum_{n\neq 0} C_n e^{-ink}$ represents the complex energy band of the frequency lattice, with $k$ being the Bloch momentum [18]. Eq. (2) reveals the core physical principle of our scheme that the flat-comb condition is exactly equivalent to the trajectory of the complex energy band $E_k$ in parametric space. This equivalence directly connects comb-envelope engineering with non-Hermitian band topology, enabling topologically steerable frequency evolution.

***Flat comb generation via nonreciprocal frequency transition***—We first consider the case of nearest-neighbor coupling. Substituting $C_1 = J - \Delta\exp(-i\varphi)$ and $C_{-1} = J + \Delta\exp(+i\varphi)$ into Eq. (2) yields the flatness condition

$$X^2 + Y^2 + 2XY\sin(\varphi) = \cos^2(\varphi), \quad (3)$$

where $X = \delta\omega/(2J)$, $Y = \gamma/(4\Delta)$ are dimensionless parameters determined by the laser detuning, cavity dissipation, and the modulation strength and phase. For the Hatano-Nelson model ($\varphi = 0$) [35], Eq. (3) reduces to $X^2 + Y^2 = 1$, describing a circle in the $(X, Y)$ parametric space. Figure 2(a) shows the theoretically calculated distribution of $|L|$ obtained from the steady-state solution of Eq. (1). The yellow curve depicting flat combs ($|L| = 1$) closely follows the circular trajectory, which separates two distinct transport regimes of $|L| < 1$ and $|L| > 1$. In the limit $Y\rightarrow\infty$ ($\Delta\rightarrow 0$), the non-Hermitian contribution vanishes and the system returns to the conventional Hermitian case.

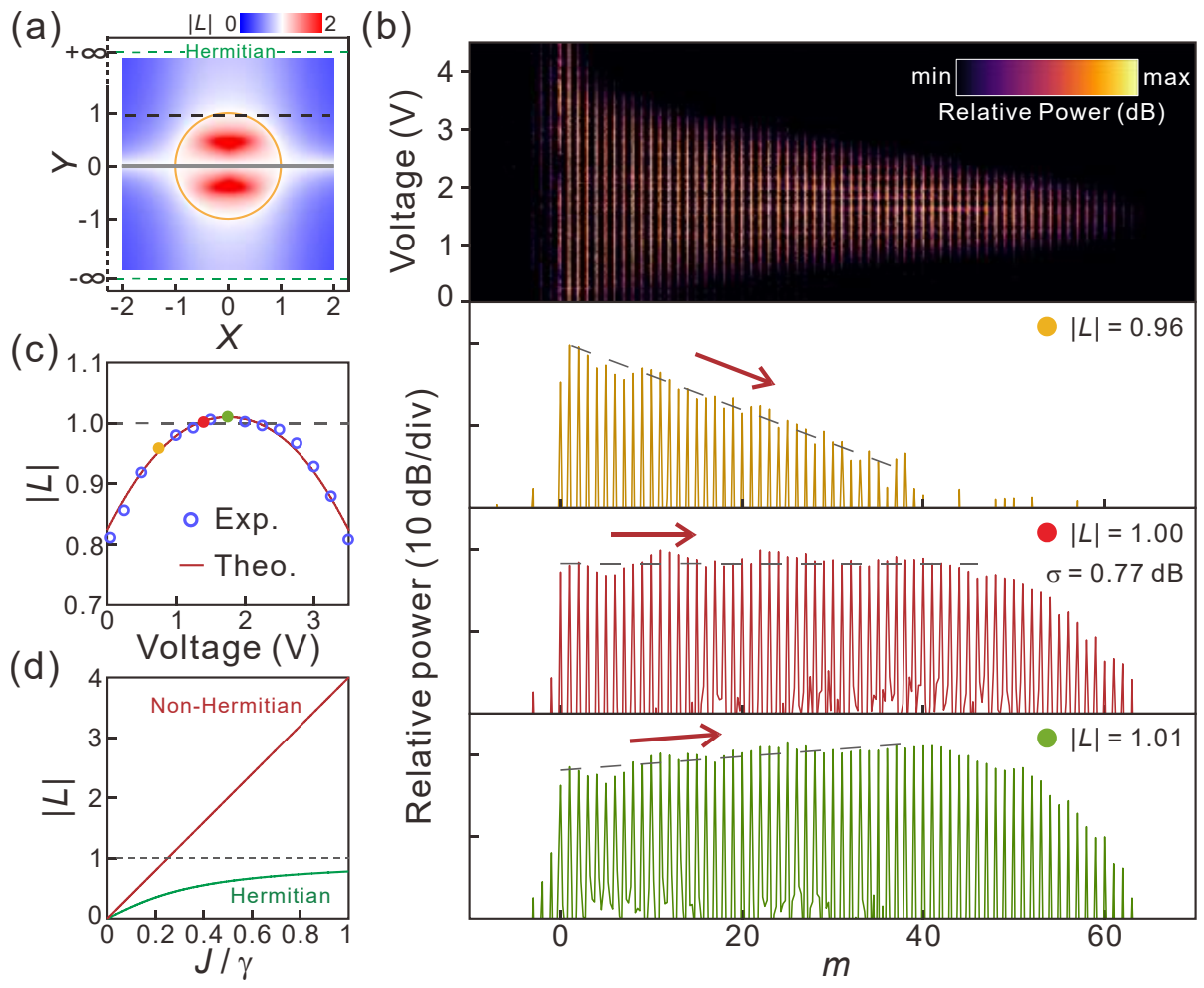

**Fig. 2. Flat EO comb generation via directional frequency evolution.** (a) Calculated phase diagram of the flat-comb parameter $|L|$. $X = \delta\omega/(2J)$, $Y = \gamma/(4\Delta)$. The yellow curve corresponds to $|L| = 1$. (b) Measured comb spectra as a function of applied bias voltage $V$ (effective detuning $\delta\omega = \Omega_R V/2V_\pi$, $V_\pi$ = 2.1 V). The three representative spectra at $|L| = 0.96$, 1.00 and 1.01 illustrate exponential decay, perfect flatness and directional spectral amplification, respectively. $\sigma$ denotes the root-mean-square deviation (RMSE) in power of the flat comb lines. (c) Extracted $|L|$ as a function of bias voltage (detuning) from the experimental data shown in (b). (d) Comparison of $|L|$ for Hermitian and non-Hermitian regimes.

For experimental demonstration, we investigate comb generation in an optical fiber loop with thin-film lithium niobate (TFLN) phase and amplitude modulators (see Supplemental Material [34]). We sweep the system across the phase diagram of Fig. 2(a) by continuously tuning the effective detuning $\delta\omega$ by applying a variable direct-current (DC) bias to an additional phase modulator. Figure 2(b) presents the measured comb spectra as a function of bias voltage. A clear directional asymmetry appears throughout the evolution, confirming the emergence of nonreciprocal frequency transitions. In particular, three representative spectra at $|L|$ = 0.96, 1.00 and 1.01 illustrate exponential decay, perfect flatness and directional spectral amplification, respectively. At the critical condition $|L| = 1$, the flat comb spans approximately 50 lines with a root-mean-square deviation (RMSE) of σ = 0.77 dB in power, demonstrating a high spectral uniformity.

In order to quantitatively characterize the evolution of the spectral envelope, we extract $|L|$ from the experimental data in Fig. 2(b) and plot the results in Fig. 2(c). As the effective detuning ($\delta\omega$) increases, $|L|$ rises above unity and then gradually descends back, consistent with the theoretical prediction in Fig. 2(a). To highlight the distinctive advantages of the non-Hermitian transport, we compare $|L|$ for Hermitian and non-Hermitian modulation regimes as a function of modulation strength, as shown in Fig. 2(d). In the Hermitian case, $|L|$ can hardly reach unity even when modulation strength approaches the practical limit, confirming that the symmetric bidirectional coupling inherent to reciprocal lattices fundamentally prevents flat comb formation. While alternative approaches based on long-range coupling and nonlinear parametric oscillation have been proposed to improve comb flatness [3,16], they impose stringent requirements on modulation frequency and strength. In contrast, the nonreciprocal coupling strategy substantially expands the accessible range of $|L|$, exceeding unity even at low modulation strengths. Therefore, our non-Hermitian approach provides a distinct mechanism for ultra-flat EO comb generation, with all cavity and modulation parameters analytically determined.

***Topological-winding-controlled directional frequency evolution***—Nonreciprocal coupling gives rise to nontrivial non-Hermitian band topology, which plays an important role in controlling frequency evolution. Here we show that the spectral flow direction is governed by the topological winding associated with the complex energy band. For a reference energy $\varepsilon \in \mathbb{C}$, the winding number is defined as [18,30,36]

$$\nu = \int_0^{2\pi} \frac{\mathrm{d}k}{2\pi i} \frac{\mathrm{d}}{\mathrm{d}k} \ln\left[E(k) - \varepsilon\right], \qquad (4)$$

which characterizes the number of times the complex energy spectrum encircles $\varepsilon$ as the Bloch momentum $k$ sweeps the Brillouin zone. In synthetic frequency lattices, the complex energy bands can be reconstructed from time-resolved transmission measurements [28],

$$T_{out}(\delta\omega, t = k) = \left|\frac{S_{out}}{S_{in}}\right|^2 = \left.\frac{\kappa^2}{\left[\gamma/2 - \mathrm{Im}(E_k)\right]^2 + \left[\delta\omega + \mathrm{Re}(E_k)\right]^2}\right|_{k=t}, \qquad (5)$$

thereby revealing the winding number in the complex plane.

In the Hermitian case ($C_1 = C_{-1}$), the band is purely real and collapses into a line segment along the real axis [(i) of Fig. 3(a)]. This band cannot encircle any reference point, yielding a trivial winding number $\nu = 0$. As a result, Bloch states with opposite group velocities coexist, producing a symmetric, bidirectional spectrum [Figs. 3(b)-3(c)]. With nonreciprocal coupling, the non-Hermitian band [(ii)-(iv) of Fig. 3(a)] acquires a finite imaginary component and forms a closed loop in the complex plane. Taking $\varepsilon = 0$ and $E(k) = -2[J\cos(\Omega_R k) + i\Delta\sin(\Omega_R k+\varphi)]$, the winding number is obtained as $\nu = \mathrm{sign}(\pi/2-|\varphi|)$. Varying $\varphi$ thus switches the winding number, which directly determines the net transport direction of the surviving Bloch states in the synthetic frequency lattice. For $\nu = -1$, Bloch states with positive group velocity dominates, driving energy toward higher frequencies [Figs. 3(d)-3(e)]. At $\varphi = \pi/2$, the winding number becomes trivial ($\nu = 0$), and the bidirectional diffusion restores, which therefore prevents the formation of a spectrally flat comb [Figs. 3(f)-3(g)]. For $v = +1$, the spectral flow direction reverses, preferentially toward lower frequencies [Figs. 3(h)-3(i)].

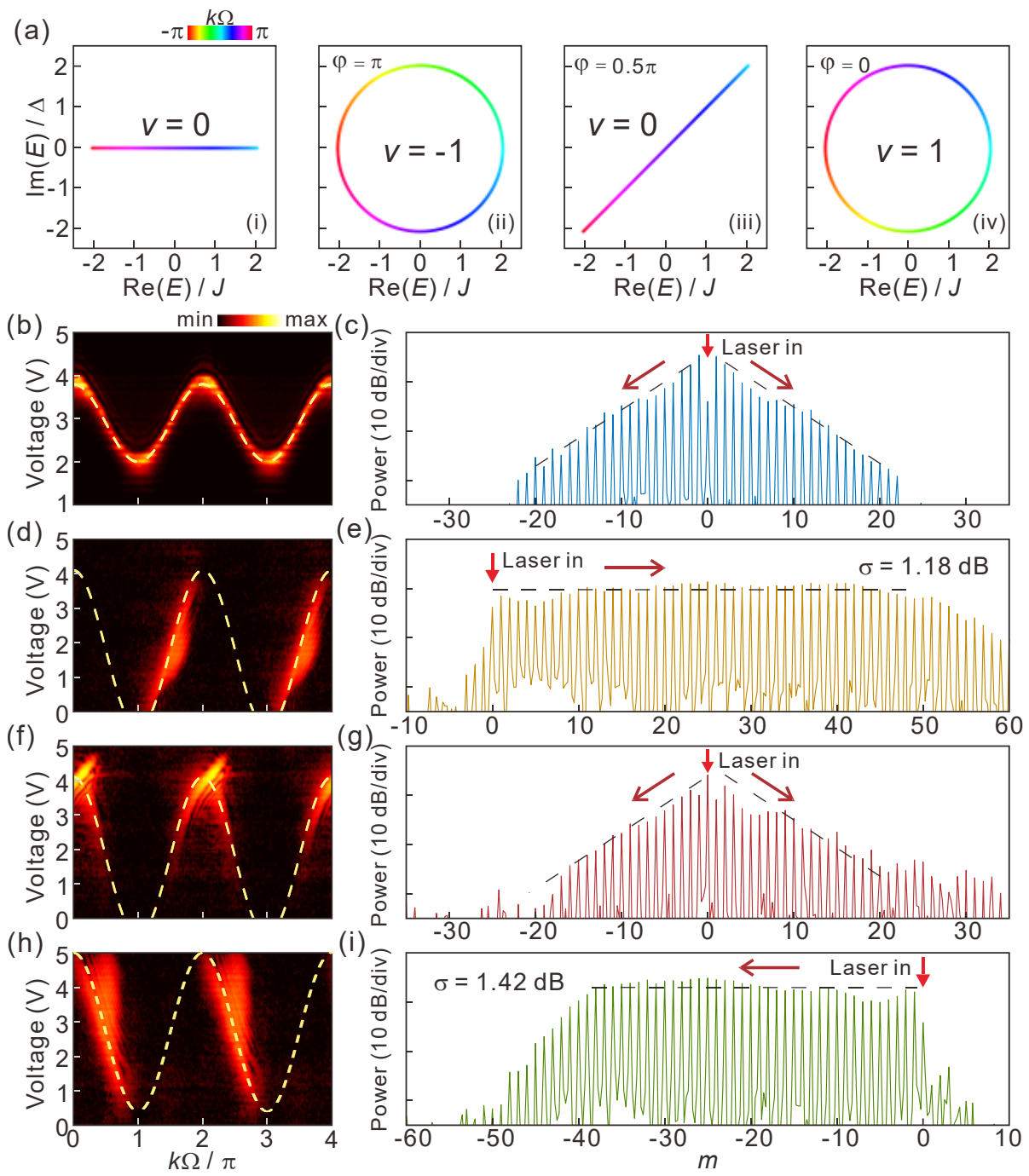

**Fig. 3. Topological-winding-controlled directional frequency evolution.** (a) Calculated energy bands for Hermitian (i) and non-Hermitian (ii-iv) frequency lattices. Measured band structure and comb spectra for (b)-(c) Hermitian case (winding number $v = 0$), (d)-(e) non-Hermitian case ($v = -1$), (f)-(g) non-Hermitian case ($v = 0$), and (h)-(i) non-Hermitian case ($v = +1$).

We next examine non-Hermitian configurations beyond the minimal Hatano-Nelson model ($\varphi \neq n\pi$). For general $\varphi$, Eq. (3) yields an elliptical flatness contour ($|L| = 1$) in the ($X$, $Y$) plane [Fig. 4(a)], indicating that flat comb generation remains accessible over a continuous range of modulation phases. In the experiment, we tune the effective detuning $\delta\omega$ and measure the comb spectra at $\varphi = 5\pi/6$, $3\pi/4$ and $2\pi/3$. Figure 4(b) plots extracted variation of $|L|$ versus detuning for each $\varphi$. Each curve exhibits two points where $|L| = 1$, corresponding to intersections with the elliptical contour. As $\varphi$ varies from $5\pi/6$ to $2\pi/3$, the elliptical contour deforms continuously, while the winding number remains $v = -1$ for $\varphi > \pi/2$. Consequently, the sign of the dominant group velocity is preserved, as shown in Figs. 4(c)-4(e). Despite substantial deformation of the flatness phase diagram and redistribution of Bloch-state populations, all measured combs exhibit identical spectral flow direction toward higher frequencies [Figs. 4(f)-4(h)]. This robust directionality originates from the globally conserved winding number of the complex energy band and persists against variations in the modulation phase.

***Generalization to complex non-Hermitian architectures***—The principle established in Eqs. (1)-(2)

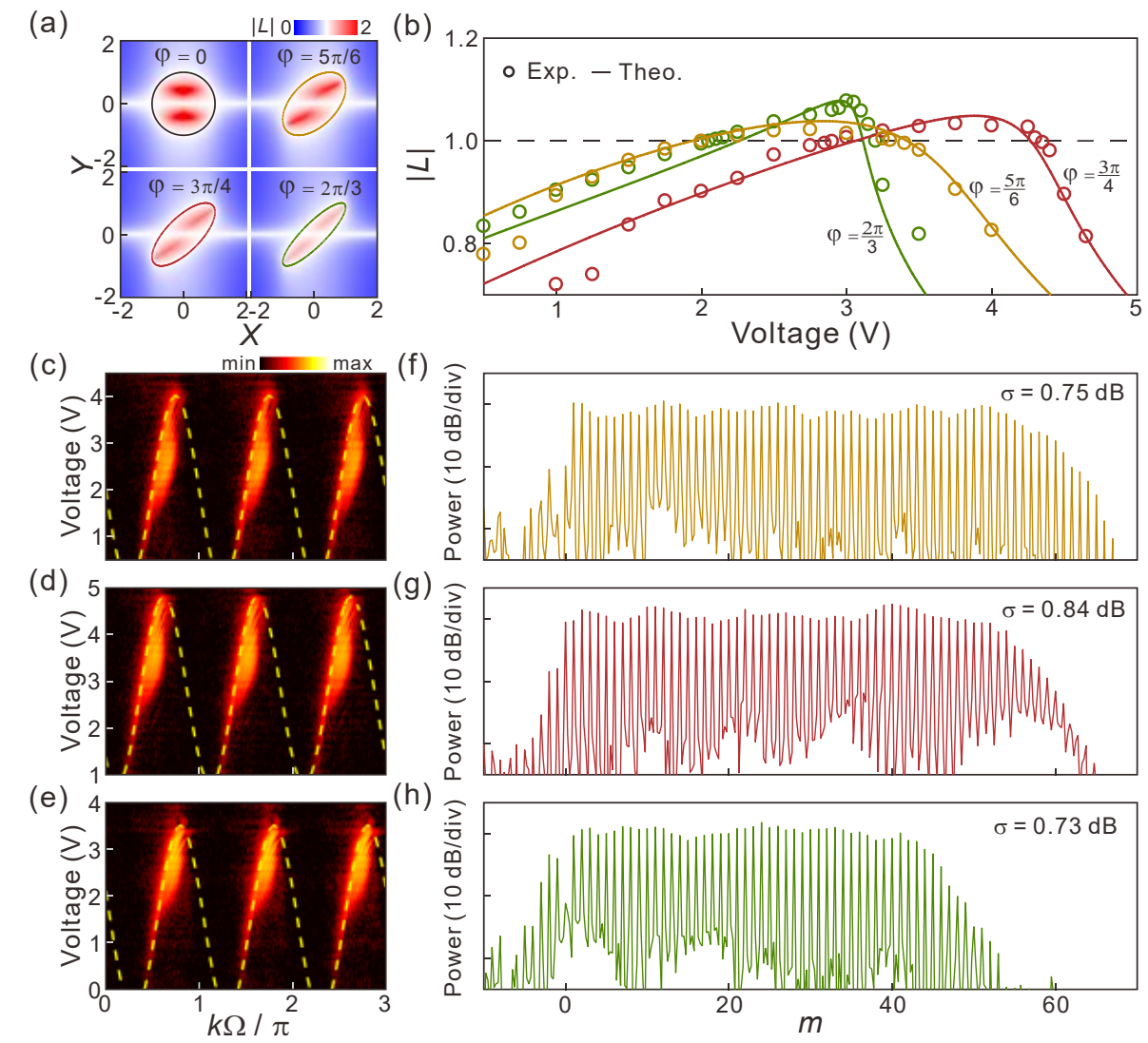

**Fig. 4. Flat EO comb generation via directional frequency evolution for general non-Hermitian configurations ($\varphi \neq n\pi$).** (a) Calculated phase diagram of the flat-comb parameter $|L|$ for $\varphi = 0$, $5\pi/6$, $3\pi/4$ and $2\pi/3$. (b) Measured (dots) and theoretical (solid curves) values of $|L|$ as a function of bias voltage (detuning) for $\varphi = 5\pi/6$, $3\pi/4$ and $2\pi/3$. (c)-(e) Measured band structure for $\varphi = 5\pi/6$, $3\pi/4$, and $2\pi/3$, respectively. (f)-(h) Corresponding comb spectra for $\varphi = 5\pi/6$, $3\pi/4$, and $2\pi/3$, respectively. All the generated combs exhibit a highly flat spectral envelope (RMSE $\sigma < 1$ dB).

generalizes naturally beyond nearest-neighbor interactions to long-range coupling regimes, where arbitrary topological winding can be generated by applying modulations at harmonics of $\Omega_R$, i.e. $n\Omega_R$ for $n$-th-order coupling [19]. As an example, we consider a second-order configuration with $C_1 = C_{-1} = J$, $C_{-2} = \Delta\exp(i\varphi)$, and $C_2 = -\Delta\exp(-i\varphi)$ in Appendix A. The corresponding flatness condition can be derived as $X = \cos(\alpha)$, $Y = \sin(2\alpha-\varphi)$. Notably, long-range coupling enables much richer topological behaviors as the winding number can be either +1 or -1, thus allowing the spectral flow direction to be switched by tuning the cavity detuning (parameter X) across different branches of the trajectory. Beyond single-lattice generalizations, the same principle can be extended to multi-lattice architectures. Appendix B presents a non-Hermitian diatomic frequency lattice, and the corresponding output exhibits a flat dual comb, which can be further scaled to synthesize flat multiple combs. Collectively, these extensions demonstrate that our non-Hermitian flat-comb scheme is not limited to nearest-neighbor coupling but represents a general principle applicable to a wide range of lattice configurations.

***Discussion and Conclusions***—While flat comb generation has been demonstrated using other approaches, including cascaded modulation, long-range coupling, and nonlinear spectral broadening [3,15,16], our non-Hermitian strategy offers distinct advantages (see Table S1, Supplemental Material [34]). First, it does not rely on strong optical nonlinearities or complex cavity structures, thus facilitating compact and energy-efficient operation. Second, within a non-Hermitian frequency lattice, perfect spectral flatness can be achieved under rigorously established conditions, providing analytical predictability for device implementation. Moreover, the nonreciprocal coupling gives rises to a topological winding that governs the spectral flow direction, rendering it robust against detuning and phase variations. These features draw from non-Hermitian band topology to establish a direct connection with flat comb generation, thereby overcoming the exponential-decay bottleneck inherent to conventional Hermitian schemes and providing a deterministic route to flat, tunable EOFCs. Although the observable flat comb lines in our current experiment is limited, primarily due to fast thermal fluctuations in the fiber cavity, theoretical analysis shows that the number of comb lines scales with the effective roundtrips of resonance evolution (see Supplemental Material [34]). A broadband flat comb can therefore be achieved by improving the stability of the modulated cavity, for example through active feedback stabilization [37] or by adopting thermally compensated chip-scale platforms [20].

In summary, we have demonstrated topologically steerable ultra-flat EO combs through engineered nonreciprocal coupling in a non-Hermitian frequency lattice. The exact flatness condition, derived from the coupled-mode dynamics, maps precisely onto the complex band trajectory, establishing a direct connection between comb-envelope engineering and non-Hermitian topology. For experimental verification, we achieve a highly uniform spectral envelope for EO comb generation, with the spectral flow direction governed by the topological winding of the non-Hermitian band. The extension to long-range coupling and multi-lattice architectures underscores the versatility of this non-Hermitian scheme as a powerful tool for spectral engineering. Furthermore, the mechanism is compatible with thin-film lithium-niobate (TFLN) integration [12,38], opening a route toward chip-scale, topologically steerable EO combs for optical communications, microwave photonics, and LiDAR.

## ACKNOWLEDGMENTS

We acknowledge helpful discussions on experimental measurements with P. St-Jean and Chi Zhang. This work was supported by the National Key Research and Development Program of China (Grant No. 2025YFA1411800, No. 2023YFA1406800), National Natural Science Foundation of China (No. 62375097, No. 12374305, No. 12021004), and the Hubei Provincial Young Top-notch Talent Program.

## End Matter

***Appendix A: Flat frequency comb generation under long-range coupling.*** Eq. (2) establish a general principle for flat EO comb condition that extends beyond nearest-neighbor interactions to long-range coupling regimes. For the frequency lattice with a second-order configuration, $C_1 = C_{-1} = J$, $C_{-2} = \Delta\exp(i\varphi)$, and $C_2 = -\Delta\exp(-i\varphi)$ [Fig. 5(a)], the corresponding flatness condition can be derived as $X = \cos(\alpha)$, $Y = \sin(2\alpha-\varphi)$. For $\varphi = 0$, the phase diagram exhibits a Lissajous-type closed trajectory [Fig. 5(b)], which indeed coincides with the corresponding complex band trajectory. The most significant distinction from the nearest-neighbor coupling case is that the complex band supports two topological windings, $\nu = \pm 1$. This implies that by simply tuning the detuning $\delta\omega$, a pair of flat combs with opposite evolution directions can be generated. This theoretical prediction can be verified by mapping the comb spectra as a function of the detuning [Fig. 5(c)], with the two flat combs displayed in Figs. 5(d) and 5(e) respectively.

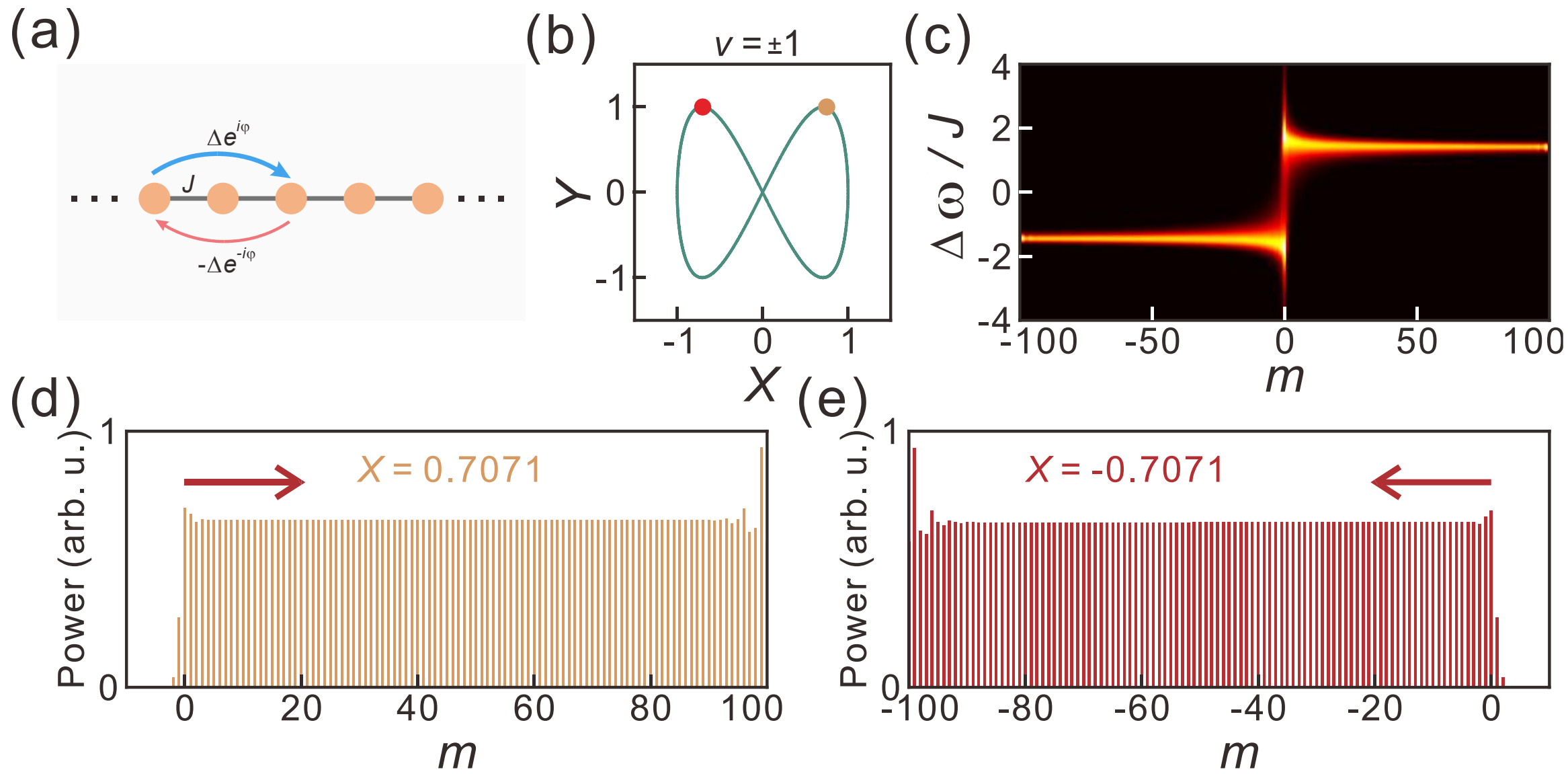

**Fig. 5. Flat frequency comb under long-range coupling.** (a) Frequency lattice with non-Hermitian long-range couplings. (b) Calculated phase diagram of the flat-comb parameter $|L|$. (c) Mapped comb spectrum as a function of detuning. (d)-(e) Flat-comb spectra with opposite evolution directions.

***Appendix B: Scaling to flat multiple frequency combs.*** The general principle established in Eq. (2) also applies to multi-atom unit cell lattice, enabling the generation of flat multiple frequency combs. We consider a system consisting of two coupled resonant rings [Fig. 6(a)], in which the symmetric and antisymmetric supermodes correspond to two lattice sites in a unit cell [19]. Non-Hermitian non-reciprocal couplings are introduced in the same manner, by applying simultaneous phase and amplitude modulations. As a result, a non-Hermitian diatomic frequency lattice can be constructed, as shown in Fig. 6(b). The coupled-mode equations describing the diatomic lattice are

$$\frac{\partial a_m}{\partial t} = -i\delta\omega a_m - \frac{\gamma}{2}a_m + iC_+ b_{m-1} + iC_- b_m + i\sqrt{\kappa}S_{in}\delta_{m,0},$$
$$\frac{\partial b_m}{\partial t} = -i\delta\omega b_m - \frac{\gamma}{2}b_m + iC_+ a_m + iC_- a_{m+1} \quad \text{(A1)}$$

where $a_m$ and $b_m$ denote the antisymmetric and symmetric modes, respectively, at the $m$-th frequency site. A single-frequency laser with amplitude $S_{in}$ excites the $a_0$ site. For nonreciprocal couplings $C_\pm = J \pm \Delta\exp(\pm i\varphi)$, the flat-comb condition derived from Eq. (3) yields $X^2 + Y^2 + 2XY\sin(\varphi) = \cos^2(\varphi)$, with the phase diagram shown in Fig. 6(c). This condition is consistent with that of the single-lattice case, as the global topological features originate from the reciprocal couplings and persist in multi-lattice systems. Under this configuration, a flat dual frequency comb can be achieved, as shown in Fig. 6(d). Similarly, flat multiple combs can be generated with multi-lattice architectures. Figures 6(e) and 6(f) show the lattice configuration and corresponding flat spectrum for a triple-lattice architecture. Overall, these extensions demonstrate that the non-Hermitian strategy offers excellent potential and compatibility for generating flat multiple combs.

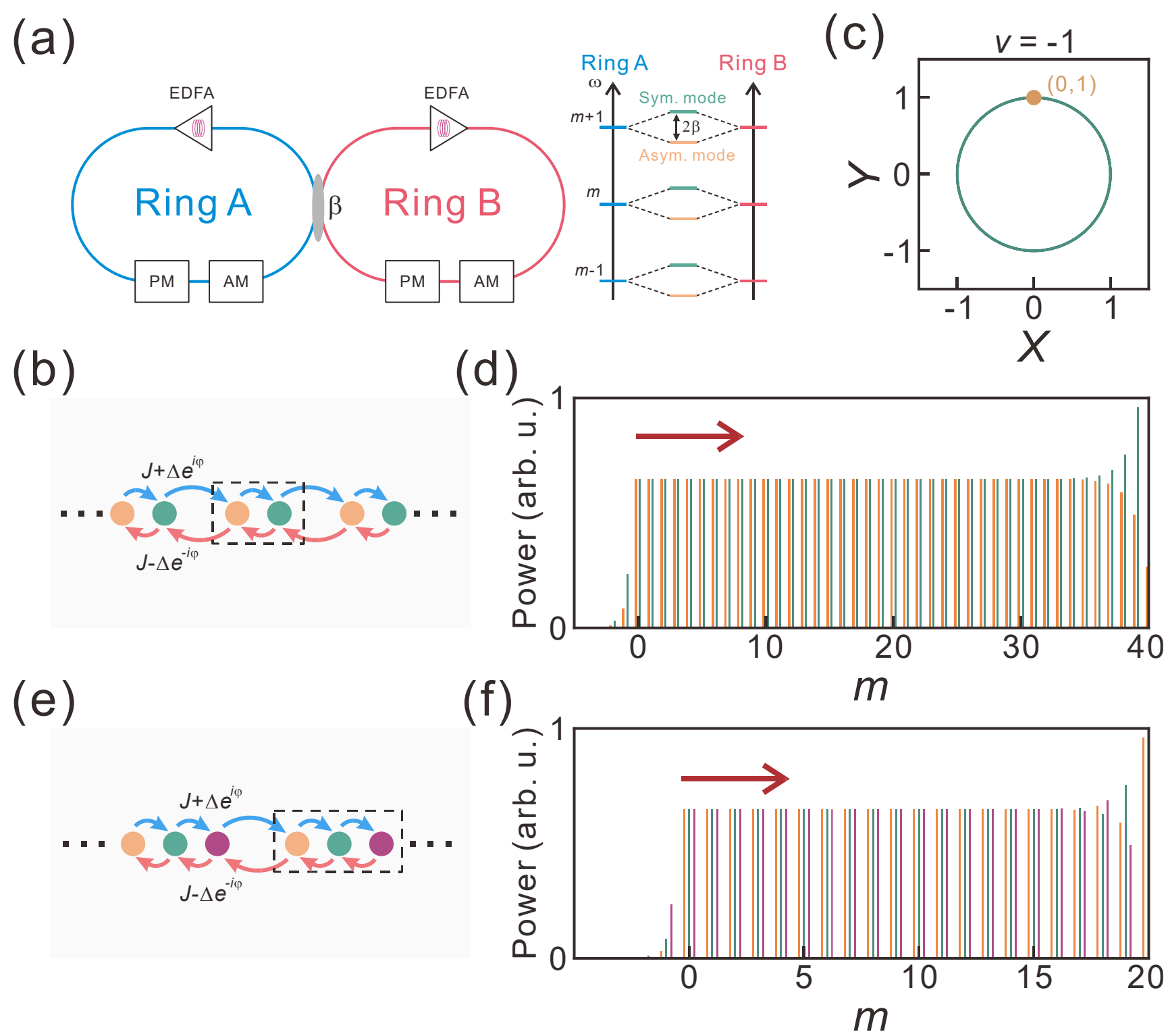

**Fig. 6. Flat dual and multiple combs with multi-lattice architectures.** (a) Two coupled resonant rings. (b) Corresponding non-Hermitian diatomic frequency lattice. (c) Calculated phase diagram of the flat-comb parameter $|L|$. (d) Calculated flat dual comb spectrum. (e) Non-Hermitian triatomic frequency lattice. (f) Corresponding flat multiple-comb spectrum.